\documentclass{article}

\usepackage{arxiv}

\usepackage[utf8]{inputenc}
\usepackage[T1]{fontenc}
\usepackage{graphicx}
\usepackage{dcolumn}
\usepackage{bm}
\usepackage{epsfig}
\usepackage{subfig}
\usepackage{placeins}
\usepackage{graphicx}
\usepackage{amsmath}
\usepackage{amssymb}
\usepackage{svg}
\graphicspath{{Fig/}}
\usepackage[labelfont=bf,justification=RaggedRight,format=plain]{caption}
\usepackage{threeparttable}
\usepackage{multirow}
\usepackage{url}
\usepackage{hyperref}       

\allowdisplaybreaks

\title{Modeling material transport regulation and traffic jam in neurons using PDE-constrained optimization}

\author{Angran Li \\
	Department of Mechanical Engineering\\
	Carnegie Mellon University\\
	Pittsburgh, PA 15213 \\
	\texttt{angranl@andrew.cmu.edu} \\
	\And
	\hspace{1mm}Yongjie Jessica Zhang* \\
	Department of Mechanical Engineering \& \\ Department of Biomedical Engineering\\
	Carnegie Mellon University\\
	Pittsburgh, PA 15213 \\
	\texttt{jessicaz@andrew.cmu.edu} \\
}

\renewcommand{\shorttitle}{Modeling material transport regulation and traffic jam in neurons using PDE-CO}

\hypersetup{
pdftitle={A template for the arxiv style},
pdfauthor={Angran Li, Yongjie Jessica Zhang},
pdfkeywords={Neuron transport, Traffic jam, Microtubule swirls,PDE-constrained optimization, Isogeometric analysis},
}

\begin{document}
\maketitle

\begin{abstract}
	The intracellular transport process plays an important role in delivering essential materials throughout branched geometries of neurons for their survival and function. Many neurodegenerative diseases have been associated with the disruption of transport. Therefore, it is essential to study how neurons control the transport process to localize materials to necessary locations. Here, we develop a novel optimization model to simulate the traffic regulation mechanism of material transport in complex geometries of neurons. The transport is controlled to avoid traffic jam of materials by minimizing a pre-defined objective function. The optimization subjects to a set of partial differential equation (PDE) constraints that describe the material transport process based on a macroscopic molecular-motor-assisted transport model of intracellular particles. The proposed PDE-constrained optimization model is solved in complex tree structures by using isogeometric analysis (IGA). Different simulation parameters are used to introduce traffic jams and study how neurons handle the transport issue. Specifically, we successfully model and explain the traffic jam caused by reduced number of microtubules (MTs) and MT swirls. In summary, our model effectively simulates the material transport process in healthy neurons and also explain the formation of a traffic jam in abnormal neurons. Our results demonstrate that both geometry and MT structure play important roles in achieving an optimal transport process in neuron.
\end{abstract}


\section*{Introduction}

The neuron exhibits a highly polarized structure that typically consists of a single long axon and multiple dendrites which are both extended from its cell body. 
Since most of the materials necessary for the neuron are synthesized in the cell body, they need to experience long-distance transport in axons or dendrites to reach their effective location \cite{segev2000untangling,swanger2013dendritic}. The intracellular material transport is therefore especially crucial to ensure necessary materials are delivered to the right locations for the development, function, and survival of neuron cells. The disruption of intracellular transport can lead to the abnormal accumulations of certain cellular material and extreme swelling of the axon, which have been observed in many neurological and neurodegenerative diseases such as Huntington’s, Parkinson’s, and Alzheimer’s disease \cite{de2008role,gunawardena2005polyglutamine, millecamps2013axonal, kononenko2017retrograde, zhang2018modulation}. Therefore, it is essential to study and understand mechanisms of the transport function and dysfunction.

Recent studies have shown that the neuron is critically dependent on molecular motors to transport various materials along the longitudinal cytoskeletal structure like microtubules (MTs) \cite{vale2003molecular,Hirokawa2010,franker2013microtubule}. MTs are long and polarized polymers with biophysically distinct plus and minus ends \cite{van2008microtubule,tsukita1981cytoskeleton,schnapp1982cytoplasmic}. The polarity of MTs can decide the preferred direction in which an individual molecular motor moves. For instance, molecular motors from the kinesin and dynein superfamilies have been identified to convey materials along MTs towards their plus and minus ends respectively \cite{hirokawa2008intracellular, may2005loss}.  
Inspired by these findings, there have been many mathematical models proposed to quantitatively study the motor-driven transport process and understand the pathology of neuron diseases. 
For instance, the partial differential equations (PDEs) of linear reaction-hyperbolic form have been used to approximate the traveling waves of a single moving species \cite{allen1982fast}. This model was further extended to account for multiple moving species \cite{reed1990approximate} and their diffusion \cite{smith2001models, friedman2005model}. Based on PDE-based transport, stochastic models have also been developed for both axonal transport \cite{brooks1999probabilistic, popovic2011stochastic} and dendritic transport \cite{newby2010quasi, bressloff2014stochastic}. In addition, several mathematical models were developed to simulate material transport in unhealthy neurons. Xue \textit{et al.} presented a stochastic model to explain the segregation of MTs and neurofilaments in neurological diseases \cite{xue2015stochastic}. Bertsch \textit{et al.} proposed to couple Smoluchowski equations and kinetic-type transport equations to study the onset and progression of Alzheimer’s disease \cite{bertsch2017alzheimer}. 

Though the aforementioned PDE and stochastic models can successfully simulate and explain certain phenomena during transport, most of these models were solved only in simple one-dimensional (1D) or 2D domains without considering the complex neuron geometry. 
Recent developments in numerical methods allow us to obtain accurate solution of PDEs in complex geometries. 
Specifically, isogeometric analysis (IGA) \cite{HUGHES20054135} directly integrates geometric modeling with numerical simulation and achieves better accuracy and robustness compared to the conventional finite element method (FEM), making it a perfect tool to tackle the highly branched neuron geometry. In particular, IGA performs simulation with different types of splines as basis functions instead of Lagrange polynomials used in conventional FEM. The same smooth spline basis functions \cite{piegl2012nurbs} used for both geometrical modeling and numerical simulation lead to accurate geometry representation with high-order continuity and superior numerical accuracy in simulation. Therefore, IGA has been extensively used in shell analysis \cite{benson2010isogeometric, casquero2017arbitrary, casquero2020seamless, wei2021analysis}, cardiovascular modeling \cite{zhang2007patient,zhang2012atlas,zhang2013challenges, urick2019review,yu2020anatomically, zhang2016geometric}, neuroscience simulation \cite{li2019isogeometric,pawar2020neuronseg_bach}, fluid-structure interaction \cite{bazilevs2006isogeometric, casquero2016hybrid, casquero2018non, casquero2021divergence}, as well as industrial application \cite{yu2020hexgen, yu2021hexdom}. Truncated T-splines \cite{wei2017truncated1, wei2017truncated2} were developed to support local refinement over unstructured quadrilateral and hexahedral meshes. Blended B-splines \cite{wei2018blended} and Catmull-Clark subdivision basis functions \cite{li2019hybrid} were investigated to enable improved or even optimal convergence rates for IGA. With the advances in IGA, we developed an IGA-based simulation platform to accurately reconstruct complex neuron geometries and solved a 3D motor-assisted transport model within them \cite{li2019isogeometric}. We also developed a deep learning framework based on the IGA simulation platform to predict the material transport process in complex neurite networks \cite{li2021deep}. The results from our IGA solver showed how the complex neuron geometry affects the spatiotemporal material distribution at neurite junctions and within different branches. However, the motor-assisted model only provides a simplified model of the actual transport process but ignores the active regulation from neuron itself. 

To model the active regulation from neurons to control the transport process, we propose to use PDE-constrained optimization (PDE-CO). PDEs are commonly used in science and engineering to mathematically represent biological and physical phenomena. Recent advances in numerical methods and high-performance computing equip the development of large-scale PDE solvers. As a result, PDE-CO problems arise in a variety of applications including optimal design \cite{borggaard1997pde, brandenburg2012advanced, hinze2008optimization}, optimal control \cite{biegler2007real, herzog2010algorithms, rees2010optimal}, and inverse problem \cite{haber2012effective, xun2013parameter}. 
In particular, PDE-CO has important biomedical applications in exploiting valuable information from real medical data. For instance, Hogea \textit{et al.} presented a PDE-CO framework for modeling gliomas growth and their mass-effect on the surrounding brain tissue \cite{hogea2008image}. Kim \textit{et al.} proposed a transport-theory-based PDE-constrained multispectral imaging algorithm to reconstruct the spatial distribution of chromophores in tissue \cite{kim2011transport}. Melani utilized the blood flow data and solved a PDE-CO problem based on fluid-structure interaction to estimate the compliance of arterial walls in vascular networks \cite{melani2013adjoint}. PDE-CO problems was also used to model tumor growth model by fitting the numerical solution with real experiment data and estimating unknown parameters in the model \cite{knopoff2013adjoint, quiroga2015adjoint}. 

In this study, we develop a novel IGA-based PDE-CO framework to simulate the material transport regulation and investigate the formation of traffic jams and swirl during the transport process in complex neurite structures. Specifically, we design a new objective function in the PDE-CO model to simulate the control mechanism to (1) mediate the transport velocity field; and (2) avoid the traffic jam caused by local material accumulation. The control strength can be adjusted through two penalty parameters in the objective function. We can also modify the governing PDEs to study the formation of traffic jam. In particular, our model can simulate the traffic jam caused by the reduction of MTs and MT swirls during transport. To further study the influence of geometry on transport, we simulate material transport in two neuron tree structures with swelling geometry. In summary, our simulation reveals that the molecular motors and MT structure play fundamental roles in controlling the delivery of material by mediating the transport velocity on MTs. The defective transport on MTs can cause material accumulation in a local region which may further lead to the degeneration of neuron cells. Combined with geometry of the neurite network, the motor-assisted transport on MTs controls the routing of material transport at junctions of neurite branches and effectively distributes transported materials throughout the networks. Therefore, our study provides key insights into how material transport in neurite networks is mediated by MTs and their complex geometry. Our IGA optimization framework is also transformative and can be extended to solve other PDE-CO models of cellular processes in complex neurite networks.

\section*{IGA-based material transport optimization in neurons}
Our interest lies in the transport of particles along an axon or dendrite in neuron cell. In our previous work, we simulated the material transport process using a macroscopic molecular-motor-assisted transport model without any transport control \cite{li2019isogeometric}. Built upon this transport model, we propose a novel transport optimization model to further study the transport control mechanism of neuron and predict the formation of a traffic jam in abnormal neurons. The proposed optimization problem is described as
\begin{subequations}
    \label{eq:PDE-CO}
    \begin{alignat}{3}
         & \text{minimize}   & \quad & \mathcal{J}(n_\pm, \bm{v}_\pm, f_\pm) = \frac{1}{2}\int_0^T \int_\Omega (\bm{v}_\pm-V_\pm)^2 d\Omega dt + \frac{\alpha}{2}\int_0^T \int_\Omega \left|| \nabla n_\pm \right||^2 d\Omega dt + \frac{\beta}{2}\int_0^T \int_\Omega f_{\pm}^{2} d\Omega dt, & \text{in $\Omega$}, \label{eq:optProb}    \\
         & \text{subject to} &       & \dfrac{\partial n_{0}}{\partial t}=-(k_{+}+k_{-})n_{0}+k'_{+} l_{+} n_{+}+k'_{-} l_{-} n_{-},                                                                                                                                                               & \text{in $\Omega$},\label{eq:constraint1} \\
         &                   &       & \dfrac{\partial (l_\pm n_{\pm})}{\partial t} + \bm{v}_{\pm} \cdot \nabla (l_\pm n_{\pm})=D_\pm \nabla^{2} (l_\pm n_{\pm}) + k_{\pm}n_{0}-k'_{\pm}(l_\pm n_{\pm}),                                                                                                             & \text{in $\Omega$},\label{eq:constraint2} \\
         &                   &       & \dfrac{\partial \bm{v}_{\pm}}{\partial t} + \bm{v}_{\pm} \cdot \nabla \bm{v}_\pm=-\nabla n_\pm + \nabla \cdot (\mu \nabla \bm{v}_\pm) + f_\pm,                                                                                                & \text{in $\Omega$},\label{eq:constraint3}\\
         &                   &       & n_0=n_i, n_{+}=\lambda_i n_i                                                                                                & \text{at inlet end},\label{eq:constraint4}\\
         &                   &       & n_0=n_o, n_{-}=\lambda_o n_o                                                                                                & \text{at outlet end},\label{eq:constraint5}
    \end{alignat}
\end{subequations}
where the open set $\Omega \subset R^d$ ($d=2$ or 3) represents the $d$-dimensional internal space of the neuron, $V_\pm$ is a predefined velocity field inside neuron; $n_0$ , $n_+$ and $n_-$ are the spatial concentrations of free, incoming (relative to the cell body; retrograde), and outgoing (anterograde) particles, respectively; $D_\pm$ is the diffusion coefficient of incoming and outgoing materials; $\bm{v}_+$ and $\bm{v}_-$ are velocities of incoming and outgoing particles, respectively; $k_{\pm}$ and $k'_{\pm}$ are rates of MT attachment and detachment of incoming and outgoing materials, respectively; $l_\pm$ represents the density of MTs used for motor-assisted transport; $f_\pm$ represents the control forces that mediate the material transport; $\mu$ is viscosity of traffic flow; $\lambda_i$, $\lambda_o$ represent the degree of loading at inlet and outlet ends, respectively \cite{smith2001models}; and $n_i$, $n_o$ represent the boundary value of $n_0$ at inlet and outlet ends, respectively. Note that in this PDE-CO model, $n_0$, $n_{\pm}$ and $\bm{v}_{\pm}$ are referred as the ``state variables'' while $f_\pm$ are referred as the ``control variables''.
In this study, we assume the MT system is unipolar that leads to a unidirectional material transport process and ignore $n_{-}$, $l_-$, $\bm{v}_-$, $k_-$, $k'_-$ terms in Eq. \ref{eq:constraint1}-\ref{eq:constraint5}. The default values of simulation parameters are summarized in Table \ref{table:parameter}.

Herein, we account for active regulation from neuron in the objective function (Eq. \ref{eq:optProb}), and we assume the optimal material transportation within neuron can be achieved by solving the proposed optimization model. The first term in Eq. \ref{eq:optProb} measures the difference between $\bm{v}_{\pm}$ and the predefined optimal velocity field $V_{\pm}$. It serves as a velocity control mechanism that neuron expects to achieve the predefined velocity field $V_{\pm}$ during transport. The second term measures the cost from concentration gradient $\nabla n_{\pm}$ within the entire neuron cell. It serves as a traffic jam control mechanism that the neuron can improve local traffic jam by detecting and avoiding high concentration gradient in the entire geometry. The value of parameter $\alpha$ represents to what extent we want to optimize the transport process and avoid traffic jams. The third term is a regularization that measures the control forces applied by neuron to mediate the transport. The value of parameter $\beta$ represents how much the neuron can affect the transport velocity. To introduce traffic jams in neurons, we modify the simulation parameters in the governing equations. In this study, we modify the spatial distribution of $l_\pm$ to model the traffic jam caused by abnormal MTs such as the reduction of MTs and MT swirls during transport.

We employ the ``all-at-once'' method \cite{stoll2013all,yilmaz2014all} and IGA to formulate and solve the optimization model (Eq. \ref{eq:optProb}) with PDE constraints (Eq. \ref{eq:constraint1}-\ref{eq:constraint5}) simultaneously. We first discretize the objective function to obtain its approximation $\mathcal{J}_h(n_\pm, \bm{v}_\pm, f_\pm)$. We also discretize PDE constraints (Eq. \ref{eq:constraint1}-\ref{eq:constraint5}) to obtain their weak form $\mathcal{B}_h(n_0, n_\pm, \bm{v}_\pm, f_\pm) =0$. Then, we build a discrete Lagrangian
\begin{equation}
\mathcal{L}_h = \mathcal{J}_h + p^T \mathcal{B}_h=0,
\end{equation}
where $p^T$ is the Lagrange multiplier and is also referred to as the ``adjoint variable''. By taking derivatives of the discrete Lagrangian with respect to state, control, and adjoint variables and setting the resulting expressions to zero, we obtain the first-order conditions, or Karush-Kuhn-Tucker (KKT) conditions. The resulting KKT system is then solved using the GMRES \cite{saad1986gmres} solver implemented in PETSc \cite{abhyankar2018petsc}. In this study, we focus on solving the proposed optimization model in 2D neuron geometries. 
\begin{figure*}[!htb]
    \centering
    \includegraphics[width = \linewidth]{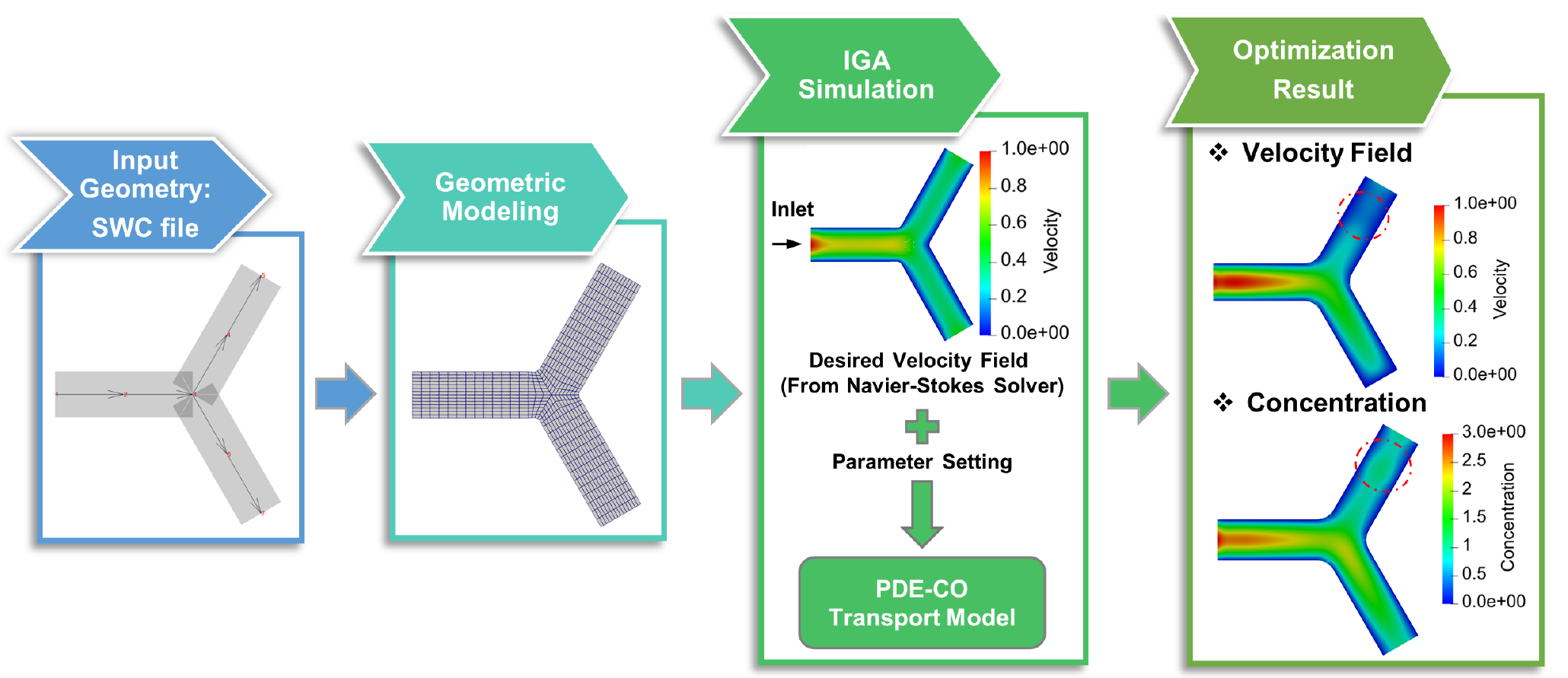}
    \vspace{-6mm}
    \caption{An overview of the material transport control simulation in a bifurcation geometry. The traffic jam is introduced by reducing MTs in the red dashed circle region. Color bars unit for velocity field: $\mu m/s$ and concentration: $mol/\mu m^3$.}
    \label{fig:Result_Bifurcation}
    \vspace{-2mm}
\end{figure*}

As shown in Fig. \ref{fig:Result_Bifurcation}, we use a bifurcation example to illustrate the pipeline of our simulation. We first generate a control mesh and reconstruct the neuron geometry with Truncated Hierarchical B-splines (THB-spline) by utilizing the geometry information stored in a SWC file. The SWC file is widely used to store neuron morphologies including vertices and the associated diameters on the skeleton of the neuron. We can obtain the SWC files for various real neuron geometries from the NeuroMorpho database \cite{ascoli2007neuromorpho}. The raw SWC file needs to be pre-processed to ensure no duplicated vertices or overlapping skeleton exist in the geometry. During the geometric modeling of our workflow, we take the cleaned-up neuron skeleton as input and use the skeleton-based sweeping method \cite{zhang2007patient} to generate quadrilateral control mesh of the neuron geometry. Then, we build THB-spline on the 
quadrilateral mesh \cite{casquero2020seamless, wei2021analysis} for the final representation of the neuron geometry. Once the spline information for the geometry is obtained, we run a steady-state Navier-Stokes solver to generate the pre-defined velocity for the optimization. We then use the default simulation parameters in Table \ref{table:parameter} and modify the spatial distribution of $l_+$ in the red circle regions to introduce traffic jam. Finally, we run the optimization solver and obtain the velocity field and concentration distribution. In this paper, we apply the pipeline to various neural structures with material transport regulation, traffic jam and MT swirl. All simulations are conducted on the XSEDE (Extreme Science and Engineering Discovery Environment) supercomputer Bridges at the Pittsburgh Supercomputer Center \cite{towns2014xsede, nystrom2015bridges}.



\section*{Results}
\subsection*{Simulation of material transport regulation and traffic jam}
\begin{figure*}[!htb]
    \centering
    \includegraphics[width = \linewidth]{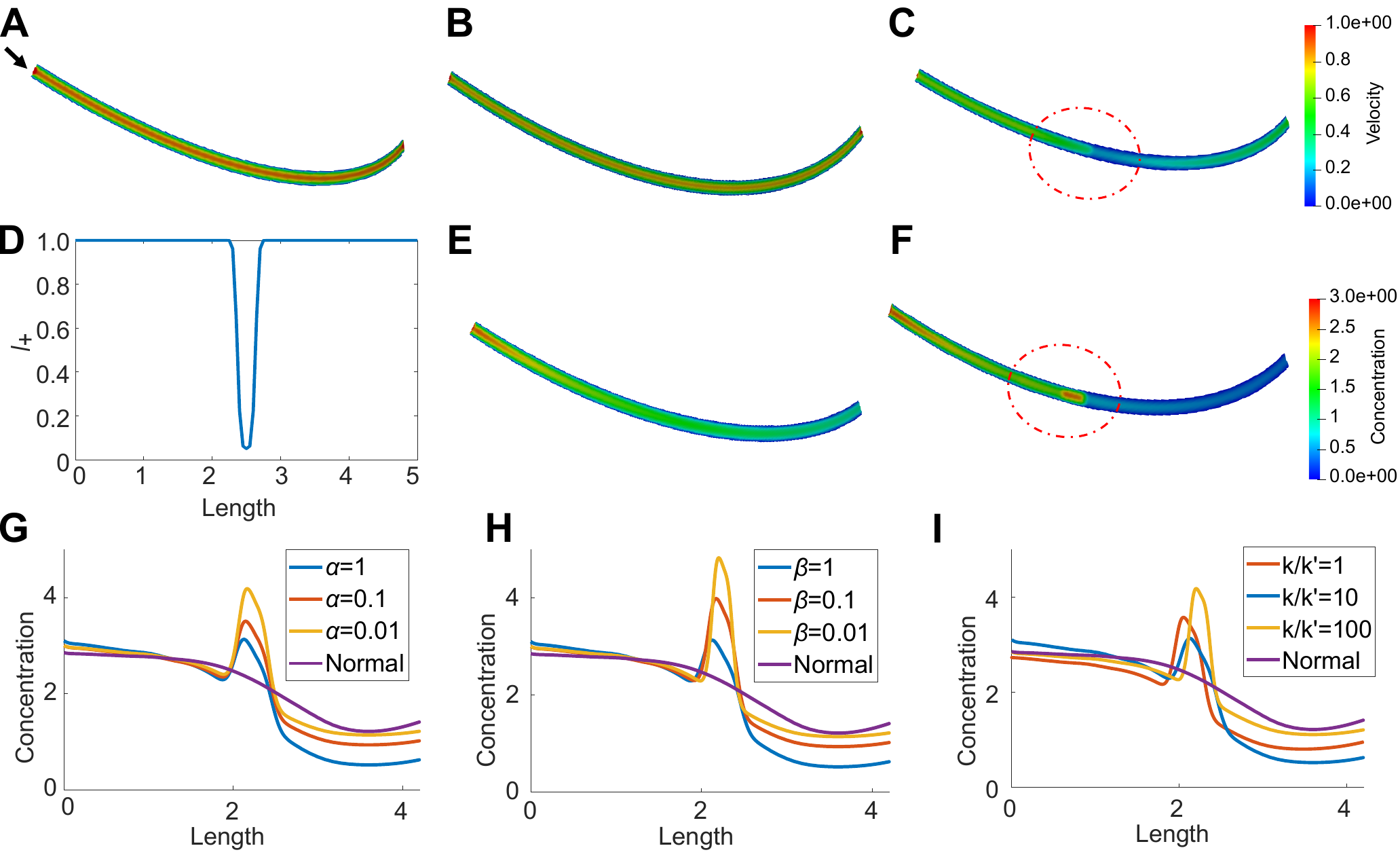}
    \vspace{-7mm}
    \caption{Simulation of material transport and parameter analysis in a single pipe geometry. (A) The predefined velocity field $V_+$. Black arrow points to the inlet of the pipe. The computed velocity field in (B) a healthy neuron and (C) an abnormal neuron with reduced MTs in the red dashed circle region. (D) Distribution of $l_+$ to model the traffic jam caused by the reduction of MTs. Distribution of concentration in (E) a healthy neuron and (F) an abnormal neuron with reduced MTs in the red dashed circle region. (G-I) The concentration curve on the centerline of the single pipe affected by different settings of (G) $\alpha$; (H) $\beta$; and (I) $k/k'$. Unit for color bars: (A-C) $\mu m/s$ and (E, F) $mol/\mu m^3$.}
    \label{fig:Result_2D_SinglePipe}
    \vspace{-2mm}
\end{figure*}

We first simulate the normal material transport and the abnormal transport with traffic jam in a single pipe geometry (Fig. \ref{fig:Result_2D_SinglePipe}). The predefined velocity field for both cases is computed by solving a steady-state Navier-Stokes equation and the result is shown in Fig. \ref{fig:Result_2D_SinglePipe}A. The other simulation parameter settings are summarized in Table \ref{table:parameter}. The computed velocity field and the distribution of concentration in the normal transport are shown in Fig. \ref{fig:Result_2D_SinglePipe}B\&E. To model traffic jam caused by the reduction of MTs, the distribution of $l_+$ along the pipe is defined as shown in Fig. \ref{fig:Result_2D_SinglePipe}D. The velocity field and material distribution results in the abnormal transport are shown in Fig. \ref{fig:Result_2D_SinglePipe}C\&F. The comparison between normal and abnormal transport shows that the velocity magnitude decreases in the red dashed circle region due to the reduced number of MTs, and this further leads to accumulation of the material in this area.
\begin{table*}
    \caption {Simulation parameters utilized in computations}
    \vspace{-4mm}
    {
        \begin{center}
            \begin{tabular}{||c l c||}
                \hline
                Parameter & Description                                                                                         & Default value \\
                \hline
                $D_\pm$   & Diffusion coefficient of incoming and outgoing materials                                            & 0.1   \\
                $k_\pm$   & Attachment rate to the MTs that transport materials in the positive (+) and negative (-) directions & 1.0 \\
                $k'_\pm$  & Detachment rate from MTs for materials that move in the positive (+) and negative (-) directions    & 0.1 \\
                $l_\pm$   & Density of MTs used for motor-assisted transport                                           & 1.0   \\
                $\mu$     & Viscosity of the traffic flow                                                                       & 0.1   \\
                $\lambda_i$     & Degree of loading at inlet end                                                                 & 2.0   \\
                $\lambda_o$     & Degree of loading at outlet end                                                                       & 2.0   \\
                $n_i$     & Boundary value of $n_0$ at inlet end                                                                      & 1.0   \\
                $n_o$    & Boundary value of $n_0$ at outlet end                                                                      & 0.0   \\
                $\alpha$    & Penalty parameter for the cost to control high concentration gradient                                                                      & 1.0   \\
                $\beta$    & Penalty parameter for the cost of control force $f_\pm$                                                                      & 1.0   \\

                \hline
            \end{tabular}
        \end{center}
    }
    \label{table:parameter}
    \vspace{-3mm}
\end{table*}

As shown in Fig. \ref{fig:Result_2D_SinglePipe}G-I and Fig. S1, we also perform parameter analysis using the single pipe geometry to study the influence of simulation parameters on the material distribution results. In particular, we focus on three parameters that may have significant effect when dealing with traffic jam caused by the reduction of MTs. The values selected for these parameters are displayed in Table \ref{table:parameterstudy}. We assume the active regulation from neuron is less dominant than natural transport via diffusion or MTs, and thus select two smaller values for $\alpha$ and $\beta$ compared to the default values in Table \ref{table:parameter}. Regarding the value selection of $k/k'$, we refer to the values utilized in \cite{kuznetsov2009macroscopic} and ensure the selected values stay within a biologically realistic range. Fig. \ref{fig:Result_2D_SinglePipe}G shows the effect of the penalty parameter of the concentration gradient cost, $\alpha$, on the concentration distribution. One can see that the decrease of $\alpha$ leads to a severer material accumulation around the region with reduced MTs in the single pipe geometry. We also find that the concentration gradient becomes larger around the traffic jam region, which indicates that there is less control over the concentration gradient due to the decrease of $\alpha$. 
\begin{table}
    \vspace{-3mm}
    \caption {Value selection for parameter study}
    \vspace{-1mm}
    {
        \begin{center}
            \begin{tabular}{||c c||}
                \hline
                Parameter                                                                                          & Value selection \\
                \hline
                $\alpha$                                                                     & 1, 0.1, 0.01\\
                $\beta$                                                                         & 1, 0.1, 0.01\\
                $k_\pm/k'_\pm$   & Fix $k_\pm=1.0\;s^{-1}$, let $k_\pm/k'_\pm$ = 1, 10, 100\\
                \hline
            \end{tabular}
        \end{center}
    }
    \label{table:parameterstudy}
    \vspace{-8mm}
\end{table}
Fig. \ref{fig:Result_2D_SinglePipe}H is similar to Fig. \ref{fig:Result_2D_SinglePipe}G but shows the effect of the penalty parameter of the control force, $\beta$, on the concentration distribution. We find similar phenomena that the traffic jam gets worse when $\beta$ decreases. 
By comparing Fig. \ref{fig:Result_2D_SinglePipe}G with \ref{fig:Result_2D_SinglePipe}H and Fig. S1A with S1B, we find $\beta$ has a greater influence on the concentration than $\alpha$ when decreasing both parameters by the same amount. 
Since $\beta$ affects the control force in Eq. \ref{eq:constraint3} while $\alpha$ affects the concentration in Eq. \ref{eq:constraint1}\&\ref{eq:constraint2}, the result indicates that the regulation of transport velocity on MTs is vital to achieve the optimal material transport process in neuron.

Fig. \ref{fig:Result_2D_SinglePipe}I shows the effect of the ratio between the attachment rate and detachment rate, $k/k'$, on the material concentration. We find that when $k/k'$ increases, the location of maximum concentration moves toward right, which indicates the decrease of detachment rate $k'$ causes more material get attached to MTs and transport faster as expected in \cite{smith2001models}. 
However, the reduction of MTs slows down the motor-assisted transport on MTs and results in worse traffic jam. Interestingly, when $k/k'$ decreases from 10 to 1, we also observe a similar traffic jam phenomena. The possible reason is that the increase of $k'$ causes more material transported via free diffusion. Although free diffusion helps to transport the material farther along the branch, the slow diffusion speed limits its ability to mitigate the traffic jam caused by the reduction of MTs.

\begin{figure*}[!htb]
    \centering
    \includegraphics[width = \linewidth]{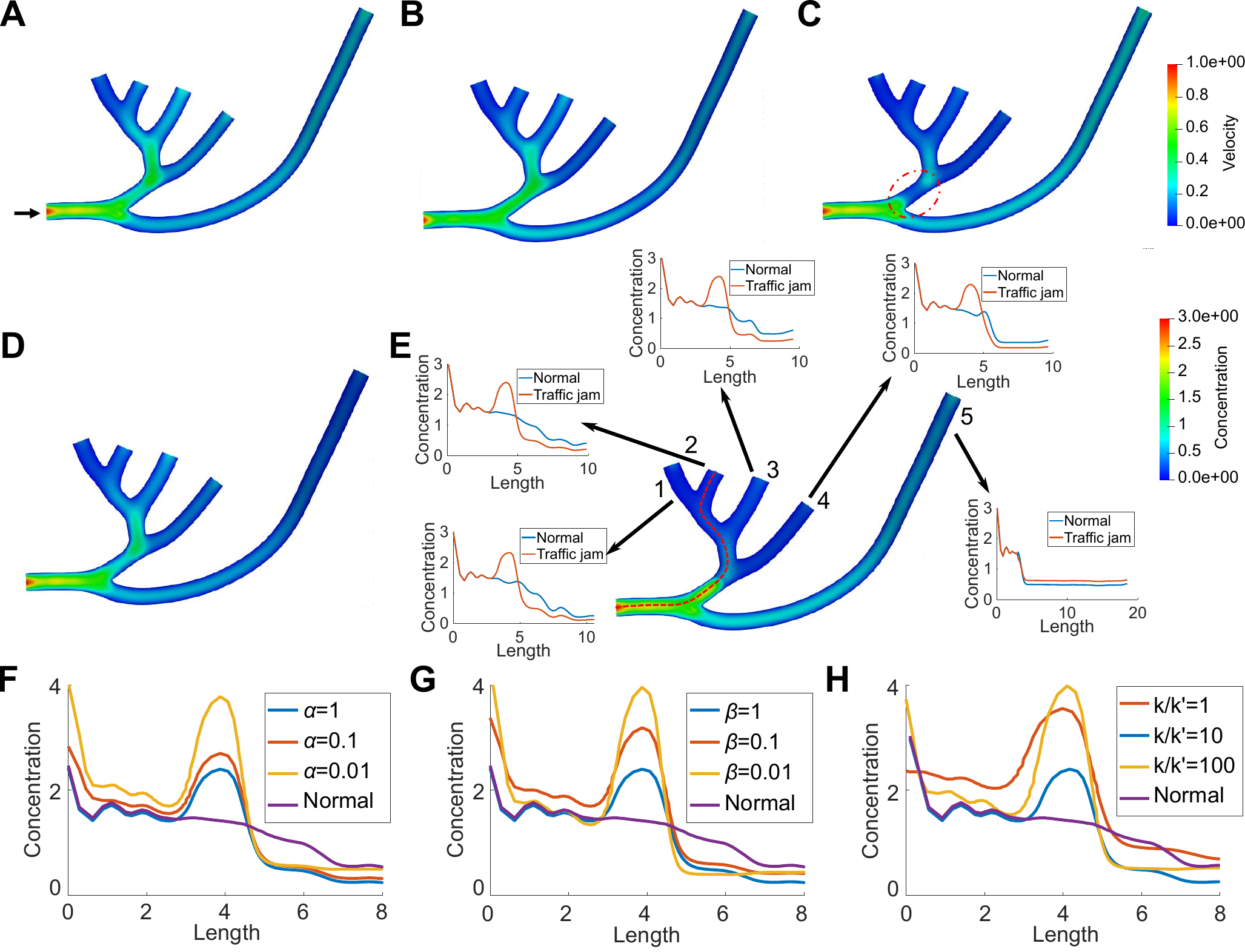}
    \vspace{-3mm}
    \caption{Simulation of material transport and parameter analysis in a neuron tree extracted from NMO\_54504. (A) The predefined velocity field $V_+$. Black arrow points to the inlet of the neuron tree. The computed velocity field in (B) a healthy neuron and (C) an abnormal neuron with reduced MTs in the red dashed circle region. Distribution of concentration in (D) a healthy neuron and (E) an abnormal neuron. We also compare the concentration curve on the centerline from the inlet to every outlet between normal and abnormal transport in (E). The red dashed curve shows the centerline from the inlet to one of the outlets and each outlet is indexed by a unique number. (F-H) The concentration curve on the centerline from inlet to outlet 2 affected by different settings of (F) $\alpha$; (G) $\beta$; and (H) $k/k'$. Unit for color bars: (A-C) $\mu m/s$ and (D, E) $mol/\mu m^3$.}
    \label{fig:Result_2D_Tree1}
    \vspace{-3mm}
\end{figure*}

\begin{figure*}[!htb]
    \centering
    \includegraphics[width = \linewidth]{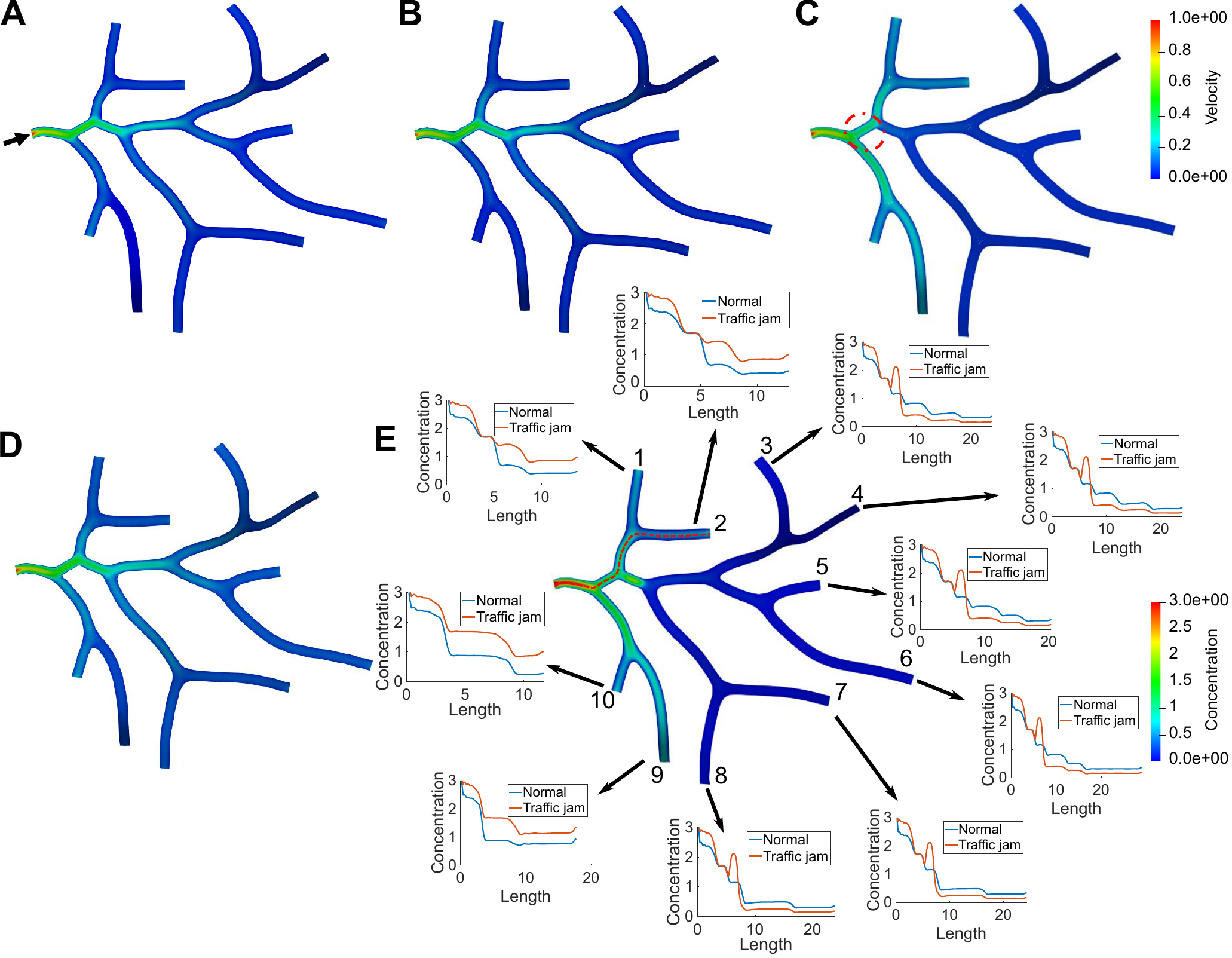}
    \vspace{-5mm}
    \caption{Simulation of material transport in a neuron tree extracted from NMO\_54499. (A) The predefined velocity field $V_+$. Black arrow points to the inlet of the material. The computed velocity field in (B) a healthy neuron and (C) an abnormal neuron with reduced MTs in the red dashed circle region. Distribution of concentration and the concentration curve on the centerline of the circled region in (D) a healthy neuron and (E) an abnormal neuron. We also compare the concentration curve on the centerline from the inlet to every outlet between normal and abnormal transport in (E). The red dashed curve shows the centerline from the inlet to one of the outlets and each outlet is indexed by a unique number. Unit for color bars: (A-C) $\mu m/s$ and (D, E) $mol/\mu m^3$.}
    \label{fig:Result_2D_Tree2}
    \vspace{-2mm}
\end{figure*}

To account for morphological effect on the transport process, we simulate the normal material transport and the abnormal transport with traffic jam in two neuron tree structures as shown in Figs. \ref{fig:Result_2D_Tree1}\&\ref{fig:Result_2D_Tree2}. The predefined velocity fields for both geometries are shown in Figs. \ref{fig:Result_2D_Tree1}A\&\ref{fig:Result_2D_Tree2}A. To quantitatively study the influence of traffic jam on the material concentration among tree structures, we also plot the concentration distribution curves along the centerline from the inlet to each outlet of these two neurons. In each curve plot, we compare the distribution between the normal transport and the abnormal transport with traffic jam, as shown in Figs. \ref{fig:Result_2D_Tree1}E\&\ref{fig:Result_2D_Tree2}E. For both cases, we model traffic jam by reducing the number of MTs ($l_+$) used for transport in the red dashed circle regions. As a result, a sudden decrease of velocity (Figs. \ref{fig:Result_2D_Tree1}C\&\ref{fig:Result_2D_Tree2}C) and material accumulation (Figs. \ref{fig:Result_2D_Tree1}E\&\ref{fig:Result_2D_Tree2}E) can be observed in these regions. 
By observing the distribution curve of the outlets downstream the traffic jam region (curve plots 1-4 of Fig. \ref{fig:Result_2D_Tree1}E and 3-8 of Fig. \ref{fig:Result_2D_Tree2}E we find that the reduced number of MTs not only causes high concentration in the local region, but also decreases the material concentration along the downstream of traffic jam region. The distribution curves of the other outlets (curve plots 5 of Fig. \ref{fig:Result_2D_Tree1}E and 1, 2, 9, 10 of Fig. \ref{fig:Result_2D_Tree2}E) demonstrate that more materials are transported to these outlets to minimize the hazard of traffic jam. The result also shows that materials rely on motor-assisted transport in longer branches of neurons and the directional transport on MTs contributes significantly to the entire transport process.

As shown in Fig. \ref{fig:Result_2D_Tree1}F-H and Fig. S2, we also perform parameter analysis on the concentration distribution in the neuron tree structure. Similar to the parameter analysis in single pipe geometry, we study the influence of three parameters on the concentration distribution and the selected values are listed in Table. \ref{table:parameterstudy}. To quantitatively study the influence, we also plot and compare the concentration curves on the centerline from inlet to outlet 2 of the neuron tree. We obtain similar results as in single pipe geometry that the decrease of $\alpha$ or $\beta$ leads to a severer material accumulation around the region with reduced MTs, and $\beta$ shows greater effect than $\alpha$ on the concentration distribution. 
In addition, we observe in Figs. \ref{fig:Result_2D_Tree1}E, \ref{fig:Result_2D_Tree1}F and S2 that when $\alpha$ or $\beta$ increases, more material is transported to the bottom long branch to mitigate the traffic jam in other branches. In Fig. \ref{fig:Result_2D_Tree1}H, we also find that the maximum concentration location moves downstream slightly when $k/k'$ increases, and either increasing or decreasing $k/k'$ intensifies the traffic jam.

\subsection*{Simulation of traffic jam with MT swirls and local swelling}
\begin{figure*}[!htb]
    \centering
    \includegraphics[width = \linewidth]{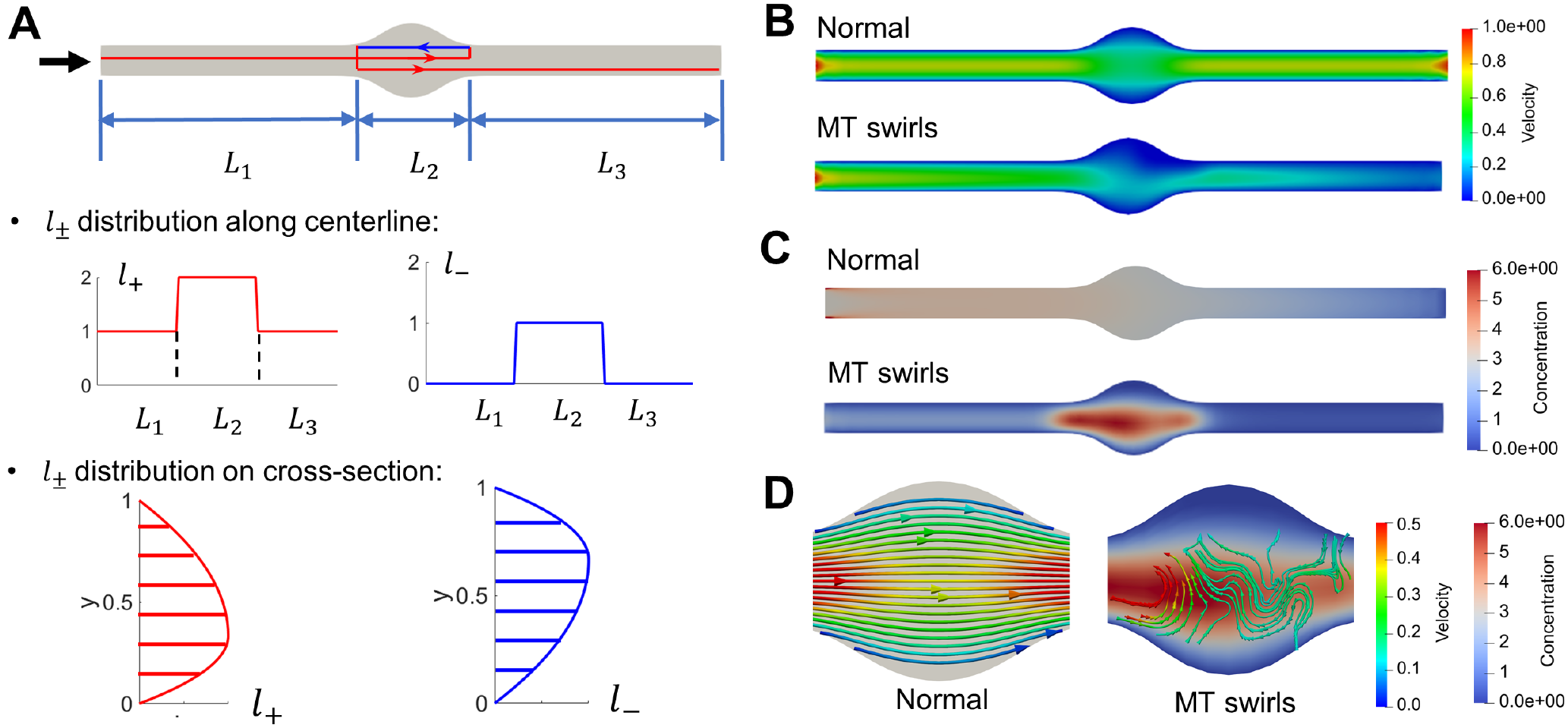}
    \vspace{-7mm}
    \caption{Simulation of material transport in a straight pipe with swelling in the middle region. (A) The simulation setting for modeling MT swirls. The red and blue arrows show the transport path along swirly MTs.  Due to the MT swirls in the $L_2$ region, both $l_+$ and $l_-$ are increased along centerline and their distributions on cross-section are also modified. (B, C) The computed velocity field and concentration distribution in the swollen geometry. (D) The velocity streamline and concentration distribution in the swollen region. Different color maps are used to distinguish between velocity and concentration. Unit for color bars: Concentration: $mol/\mu m^3$; Velocity: $\mu m/s$.}
    \label{fig:Result_2D_Pipe_swell}
    \vspace{-4mm}
\end{figure*}

Recent studies have shown that the formation of MT swirls can lead to accumulation of transported material and cause local swelling of neuron geometries \cite{shemesh2008tau}. In our model, we modify the spatial distribution of $l_\pm$ and enlarge the radius of neuron in a local region to simulate the effect of MT swirls and local swelling on the transport process. We explain the simulation setting by using a straight pipe geometry with MT swirls and swelling in the middle $L_2$ region, as shown in Fig. \ref{fig:Result_2D_Pipe_swell}. We assume the normal transport is unidirectional from left to right ($+$ direction, red arrow in Fig. \ref{fig:Result_2D_Pipe_swell}A). Due to the MT swirls in the middle region, the transport path is extended by two segments: one segment reverses to transport the material from right to left ($-$ direction, blue arrow in Fig. \ref{fig:Result_2D_Pipe_swell}A) and the other segment transports in the normal direction from left to right. Therefore, we increase the values of $l_+$ and $l_-$ along the longitudinal direction in the swelling region to describe the transport path change caused by swirling. We also assume that the swirl direction is counter-clockwise and assign an asymmetric distribution of $l_\pm$ on the cross-section in the swelling region. In particular, $l_+$ is higher on the bottom of cross-section while $l_-$ is higher on the top of cross-section, as shown in Fig. \ref{fig:Result_2D_Pipe_swell}A. We perform simulation with the new parameter setting and compare with the results of normal transport in the same geometry.
The velocity field and concentration distribution of normal and abnormal transport are compared in Fig. \ref{fig:Result_2D_Pipe_swell}B\&C, respectively. The decrease of velocity and material accumulation can be observed in the swollen region. We also find that the velocity magnitude is not symmetric anymore due to the MT swirls in abnormal transport. In Fig. \ref{fig:Result_2D_Pipe_swell}D, we plot the velocity streamline with concentration distribution in the zoomed-in swollen region for both normal and abnormal transport. Compared to the uniform velocity streamline in normal transport, the velocity displays vortex pattern in the abnormal transport, which reflects a longer transport distance due to MT swirls. We also find that the swirl of velocity streamline usually happens in the high concentration region, which implies that the material accumulation is caused by the vortex-shape velocity field.

\begin{figure*}[!htb]
    \centering
    \includegraphics[width = \linewidth]{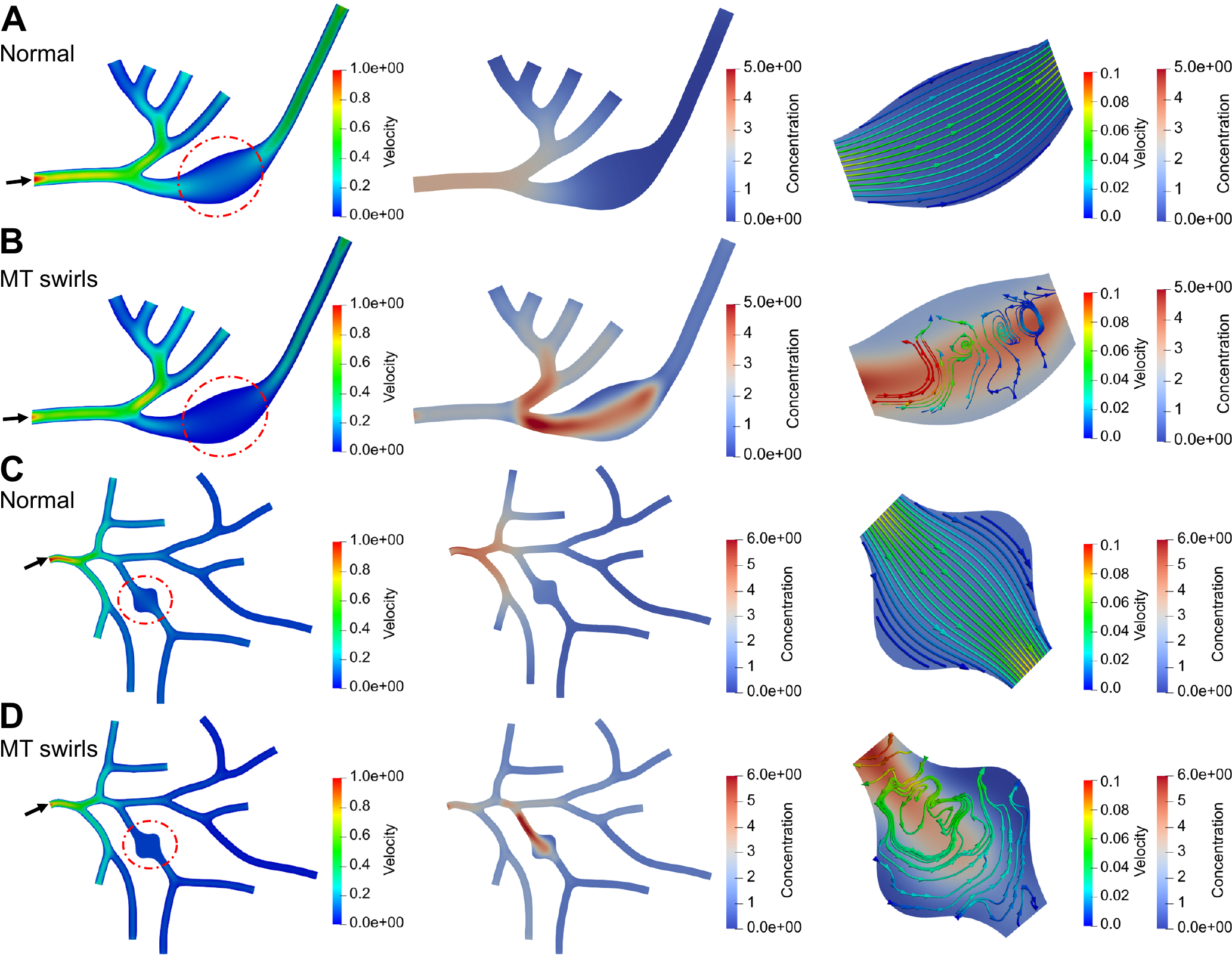}
    \vspace{-7mm}
    \caption{Simulation of material transport in neuron trees extracted from (A, B) NMO\_54504 and (C, D) NMO\_54499 with swelling in the red circle regions. The first column shows the computed velocity field and black arrow points to the inlet of the material. The second column shows the concentration distribution. The last column shows the velocity streamline and concentration distribution in the swollen region. Different color maps are used to distinguish between velocity and concentration. Unit for color bars: Concentration: $mol/\mu m^3$; Velocity: $\mu m/s$.}
    \label{fig:Result_2D_Tree1and2_swell}
    \vspace{-4mm}
\end{figure*}

As shown in Fig. \ref{fig:Result_2D_Tree1and2_swell}, we then apply the same approach to simulate the normal and abnormal transport with MT swirls in two neuron tree structures with local swelling. The swelling is introduced by increasing the skeleton radius in the red dashed circle regions. We also assume a counter-clockwise MT swirl in these swollen regions and modify the distribution of $l_\pm$ accordingly. For each model, we simulate the abnormal transport process due to MT swirls to obtain velocity field and concentration distribution results and compare with the result of normal transport in the same geometry. By comparing Fig. \ref{fig:Result_2D_Tree1and2_swell}A\&C with Fig. \ref{fig:Result_2D_Tree1and2_swell}B\&D, we find the velocity magnitude decreases and material accumulates in the swollen region. In other branches that are not downstream the swollen region, the material concentration also increases to mitigate the traffic jam in the swollen region.
In addition, similar to the results in straight pipe with swelling geometry (Fig. \ref{fig:Result_2D_Pipe_swell}D), we also observe that the velocity streamline with vortex pattern matches with the high concentration region (Fig. \ref{fig:Result_2D_Tree1and2_swell}B\&D). These results illustrate that the MT swirls lead to the circular transport velocity field in a local region which not only extends the transport distance but also traps the material and causes traffic jam.

\section*{Discussion}
\label{section:Discussion}
In this paper, we develop a PDE-constrained optimization model to simulate material transport control in neurons. Using our simulation, we examine both normal and abnormal transport processes in different geometries and discover several spatial patterns of the transport process. Our results show the formation of traffic jams due to the reduction of MTs and MT swirls in the local region. We also observe how the traffic jam affects the spatial patterns of transport velocities that in turn drives the transported materials distributed distinctly in different regions of neurite networks to mitigate traffic jam. By solving the proposed new optimization problem, we build a more realistic transport model for neurons by including active traffic regulation. The model is successfully applied to complex 2D neuron geometries and provides key insights into how neuron mediates the material transport inside its complex geometry.

Our study shows that MTs have a major impact on the material transport velocity and further affect the material concentration distribution. As shown in Fig. \ref{fig:Result_2D_SinglePipe}, the reduction of MTs in the middle of the single pipe slows down the transport velocity downstream and leads to traffic jam in the middle region. When the neuron has more branches in its geometry (Figs. \ref{fig:Result_2D_Tree1}\&\ref{fig:Result_2D_Tree2}), the reduction of MTs in one branch has a similar influence on the transport downstream the branch. However, we observe an increase in transport velocity and material concentration in other branches, indicating that the active regulation from neuron takes effect to avoid traffic jams. In addition, we perform parameter analysis to study the influence of different simulation parameters on the material concentration distribution. The ratio between the attachment rate $k$ and detachment rate $k'$ affects the amount of material transported via MTs or free diffusion. This will affect the overall transport speed and material distribution due to the different transport behaviour between motor-assisted transport and free diffusion. The penalty parameters $\alpha$ and $\beta$ affect the ability of neuron to handle traffic jams. $\beta$ has a greater influence on the traffic regulation compared to $\alpha$ since it directly affects the transport velocity on MTs, this again verifies the vital role of MTs during the intracellular transport process. Our model can also model the influence of diverse neuron topologies on material distribution. For the transport in healthy neurons (Figs. \ref{fig:Result_2D_Tree1}-\ref{fig:Result_2D_Tree2}B\&D), the magnitude of transport velocity is different among branches due to the asymmetric geometry. The different velocity magnitude further contributes to the distinct material concentration in different branches. In particular, we find that shorter branches tend to have faster transport speed and higher material concentration, which may result from the high demand of materials for their growth. 

Our study also successfully simulates and provides reasonable explanation on the traffic jam caused by MT swirls. We assume the counter-clockwise MT swirls exist in a local region of neuron geometry which cause traffic jam and geometry swelling. The spatial distribution of MT density ($l_\pm$) and neuron geometry are modified accordingly to model this phenomena. We compare the simulation result of abnormal transport on swirly MTs with normal transport and find that MT swirls have severe impact on the transport velocity field. Compared to the uniform velocity streamline in normal transport, the abnormal transport exhibits a streamline with counter-clockwise vortex pattern (Figs. \ref{fig:Result_2D_Pipe_swell}D, \ref{fig:Result_2D_Tree1and2_swell}B\&D), which is caused by the counter-clockwise MT swirls. This circular streamline not only extends the transport distance but also traps the material in the local region, and therefore explains why high concentration region matches with the circular streamline pattern.

Our study develops an IGA solver (available at \url{https://github.com/truthlive/NeuronTransportOptimization}) for solving the PDE-CO problem in complex neuron geometries. Specifically, we adopt the skeleton-based sweeping method \cite{zhang2007patient,li2019isogeometric} for mesh generation to represent the tree structures of neuron geometry. Given the geometry information of neurons, our method automatically reconstructs 2D network geometry with high accuracy and high order of continuity for IGA computation. Our automatic IGA optimization solver provides an efficient computation tool for studies of material transport regulation in complex neurite networks. The current 2D solver can be easily generalized to 3D and it is also extensible to solve other PDE-CO models of cellular processes in complex neurite network geometry.

Our study has its limitations, which we are addressing in the ongoing work. In the current model, we only consider the influence of traffic jams on the material concentration but neglect its effect on the deformation of neuron geometries. In addition, although IGA offers great advantages in accurately simulating material transport control in complex neuron geometries, the computational cost of simulating transport in large-scale neurite networks remains very expensive, which limits its biomedical application.
To improve the computational efficiency of our model, we will adopt deep learning techniques to build fast and accurate surrogate models \cite{li2020reaction, li2021deep}.
 Despite these limitations, our simulation directly shows how the traffic jam is formed in neurons and how neurons could control material traffic to avoid traffic jams. The simulation results provide references to further answer the question of how neurons deliver the right material to the right destination in a balanced manner in their complex neurite networks and how the transport may be affected by disease conditions.

\section*{Code and data availability} The source code for our model and all input data are available for download from a public software repository located at \url{https://github.com/truthlive/NeuronTransportOptimization}. All data generated during this study can be reconstructed by running the source code.

\FloatBarrier

\section*{Acknowledgememts}
The authors acknowledge the support of NSF grants CMMI-1953323 and CBET-1804929. This work used the Extreme Science and Engineering Discovery Environment (XSEDE), which is supported by National Science Foundation grant number ACI-1548562. Specifically, it used the Bridges-2 system, which is supported by NSF award number ACI-1928147, at the Pittsburgh Supercomputing Center (PSC).

\bibliographystyle{unsrt}
\bibliography{reference} 

\newpage
\newcommand{\beginsupplement}{%
        \setcounter{table}{0}
        \renewcommand{\thetable}{S\arabic{table}}%
        \setcounter{figure}{0}
        \renewcommand{\thefigure}{S\arabic{figure}}%
     }
\beginsupplement

\section*{Supplementary Information}
\begin{figure*}[!htb]
    \centering
    \includegraphics[width = \linewidth]{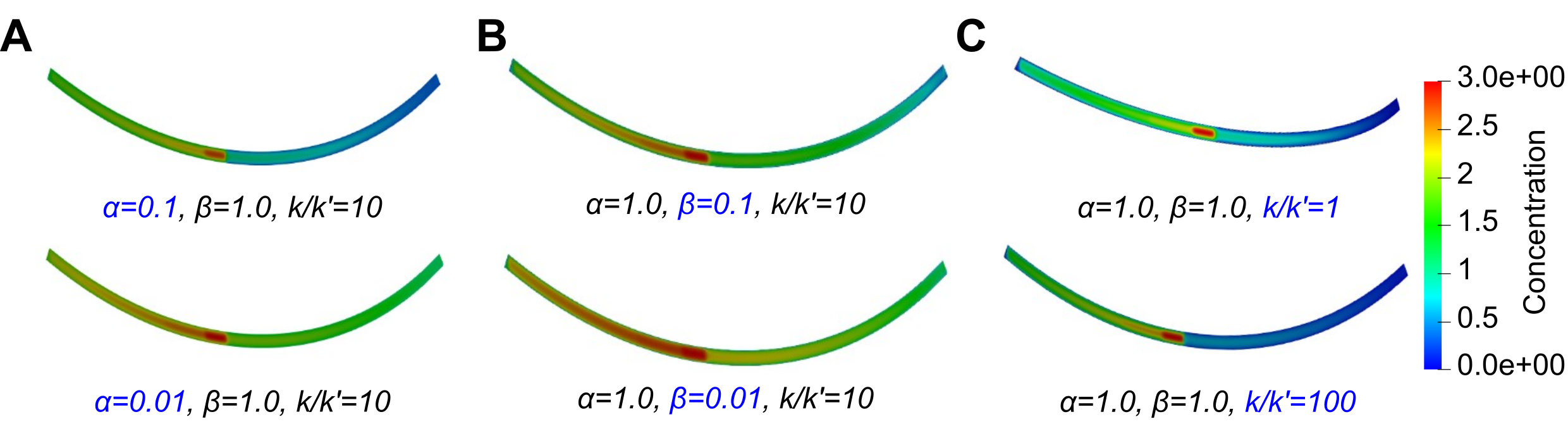}
    \caption{Parameter analysis in the single pipe geometry. The distribution of concentration affected by different settings of (A) penalty parameter $\alpha$ for the cost to control high concentration gradient; (B) penalty parameter $\beta$ for the cost of control force; and (C) the ratio between attachment rate $k$ and detachment rate $k'$. Unit for color bars: $mol/\mu m^3$.}
\end{figure*}
\begin{figure*}[!htb]
    \centering
    \includegraphics[width = \linewidth]{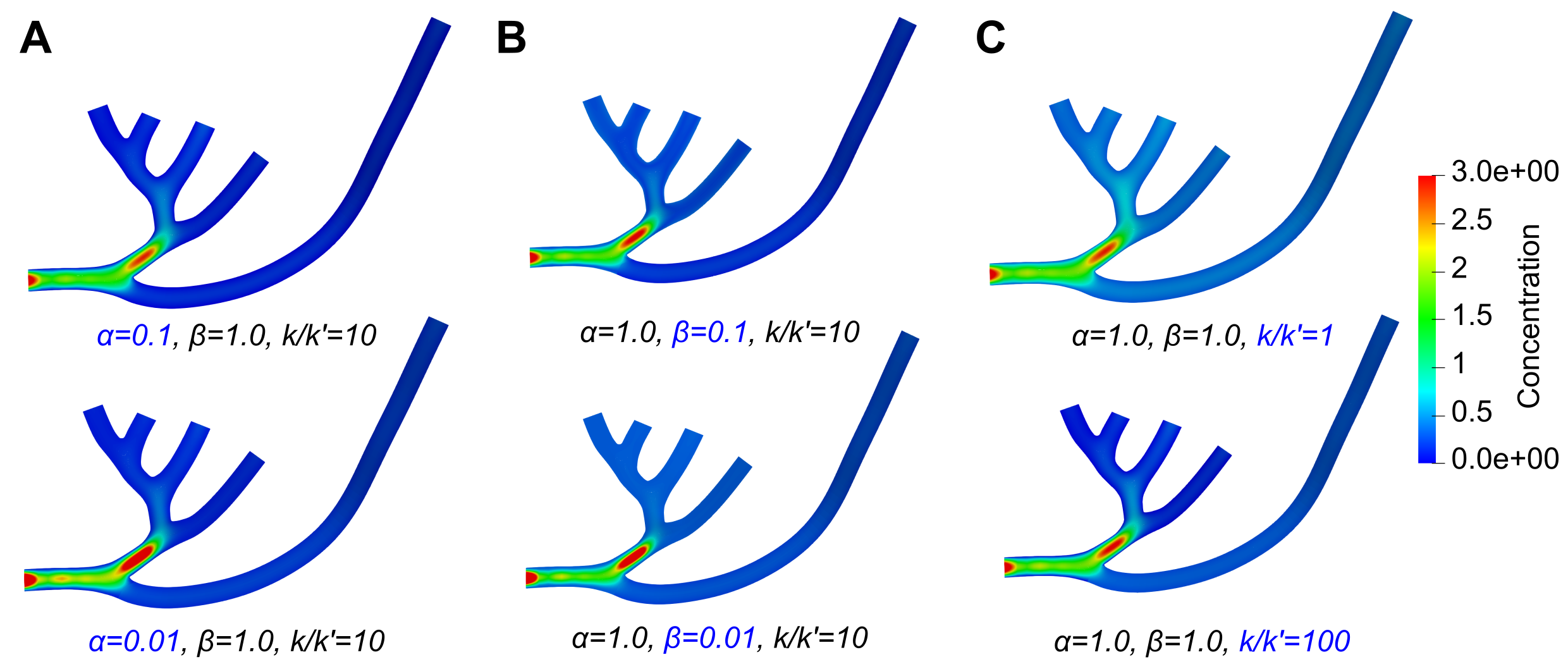}
    \caption{Parameter analysis in the neuron tree extracted from NMO\_54504. The distribution of concentration affected by different settings of (A) penalty parameter $\alpha$ for the cost to control high concentration gradient; (B) penalty parameter $\beta$ for the cost of control force; and (C) the ratio between attachment rate $k$
    and detachment rate $k'$. Unit for color bars: $mol/\mu m^3$.}
    \vspace{-4mm}
\end{figure*}
\end{document}